\documentclass[aps,twocolumn,nofootinbib,showkeys,superscriptaddress]{revtex4-2}

\usepackage{newtxtext}
\usepackage{newtxmath}
\usepackage{mathrsfs}

\usepackage{graphicx}
\usepackage{bm}
\usepackage{amsmath}
\usepackage{float}
\usepackage{multirow}
\usepackage{slashed}
\usepackage{xcolor}
\usepackage{mathtools,braket}
\usepackage{color,physics,tensor}

\usepackage[colorlinks=true, pdfstartview=FitV, bookmarks=true, bookmarksnumbered=true, breaklinks]{hyperref}
\hypersetup{linkcolor=blue, citecolor=blue, urlcolor=blue}
\newcommand{\hrefr}[2]{\href{#2}{#1}}

\begin{document}

\title{\boldmath Radiative proton capture on $\nuclide[15]{N}$ within effective field theory}

\author{Sangyeong~Son}
\email{thstkd3754@gmail.com}
\affiliation{Department of Physics, Kyungpook National University, Daegu 41566, Korea}

\author{Shung-Ichi~Ando}
\email{sando@sunmoon.ac.kr}
\affiliation{Department of Display and Semiconductor Engineering, Sunmoon University, Asan, Chungnam 31460, Korea}

\author{Yongseok~Oh}
\email{yohphy@knu.ac.kr}
\affiliation{Department of Physics, Kyungpook National University, Daegu 41566, Korea}
\affiliation{Asia Pacific Center for Theoretical Physics, Pohang, Gyeongbuk 37673, Korea}


\begin{abstract}
The astrophysical $S$ factor for the radiative proton capture process on the \nuclide[15]{N} nucleus, i.e., $\nuclide[15]{N}(p, \gamma)\nuclide[16]{O}$, 
at stellar energies are studied within the framework of the cluster effective field theory. 
The thermonuclear $\nuclide[15]{N}(p, \gamma)\nuclide[16]{O}$ reaction links the type-I to type-II cycles of the carbon-nitrogen-oxygen cycle and 
affects the abundances of elements in the univere. 
For investigating this reaction in the effective field theory formalism, we first construct an effective Lagrangian that is appropriate for this reaction 
at low-energies.
Since the intermediate excited states of the \nuclide[16]{O} nucleus have a crucial role in this reaction, we include these resonances in the formalism.
The corresponding radiative capture amplitudes and cross section are calculated, which lead to the astrophysical $S$ factor.
The low energy constants introduced in the effective Lagrangian are determined by fitting the theoretical results to the observed $S$ factors in 
the  range of $130~\mathrm{keV} < E_p < 2500~\mathrm{keV}$ using three different experimental data sets. 
Considering the recent data sets, we obtain $S(0) = 29.8\mbox{--}34.1~\mathrm{keV \ b}$, which is in a good agreement with the estimates from 
$R$-matrix approaches in the literature.
The values of $S$ at the Gamow energy are found to be larger than $S(0)$ values by about 10\%. 
\end{abstract}

\maketitle


\section{\label{sec:level1}Introduction}

Thermonuclear fusion is the energy source of stars. 
Among thermonuclear fusion processes, the proton-proton chain and the carbon-nitrogen-oxygen (CNO) cycle
are two major processes~\cite{BBFH57,AGHS10}.
The neutrinos produced in the Sun by these processes have been observed in underground 
experiments~\cite{Borexino11,Borexino18,Borexino20,GZBS21}, which brings information on the solar neutrinos and confirmed the
thermonuclear reactions of the proton-proton chain and the CNO cycle inside the Sun.
In the CNO cycle, the proton capture reaction on \nuclide[15]{N} allows two possible channels, namely, $\nuclide[15]{N} (p, \alpha) \nuclide[12]{C}$ 
and $\nuclide[15]{N} (p, \gamma) \nuclide[16]{O}$.
The former is responsible for the type-I cycle (CN cycle) and the latter for the type-II cycle (NO cycle).
At low energies the CN cycle is dominant but at higher temperatures the NO cycle becomes active.
Since the CN and NO cycles intersect at the nucleus \nuclide[15]{N}, the relative strength of the two reactions, 
$\nuclide[15]{N} (p, \alpha) \nuclide[12]{C}$ and $\nuclide[15]{N} (p, \gamma) \nuclide[16]{O}$, determines the probability of the path 
in the CNO cycle.
It turns out that the rate for the former reaction is much larger than for the latter, which means that the CN cycle is more probable than the NO cycle.
Although its contribution to the energy production in stars would be small, the $\nuclide[15]{N} (p, \gamma) \nuclide[16]{O}$ reaction is crucial
to understand nucleosynthesis and the observed oxygen abundances.
Therefore, the evaluation of its cross sections at stellar energy scales is strongly required for resolving such issues~\cite{WGUIP10,BBCC16}.

Because of the Coulomb barrier, the nuclear reaction cross section $\sigma(E)$ at the center-of-mass (c.m.) energy $E$ is parameterized 
by the astrophysical $S$ factor defined as
\begin{equation}
S(E) = E \sigma(E)  \exp(2\pi\eta),
\label{eq;S(0)}
\end{equation}
where $\eta$ is the dimensionless Sommerfeld parameter,
\begin{equation}
\eta = Z_A Z_B \mu \alpha_{\rm em} / p
\end{equation} 
with the fine structure constant $\alpha_{\rm em} = e^2/(4\pi)$.
Here, $\mu$ is the reduced mass of the system, $\mu \equiv m_A m_B / (m_A + m_B)$, where $m_A$ and $m_B$ are the masses 
of the initial state nuclei whose charges are $Z_A$ and $Z_B$, respectively.
The magnitude of the relative spatial momentum between the two nuclei in the c.m. frame is represented by $p$.

Thus the astrophysical $S$ factor is the main characteristic of any thermonuclear reaction at low energies.
Determination of its value by experiments is, however, highly nontrivial as the most experiments are carried out at the energies above 100~keV,
while the realistic energy scale for astrophysical environments is about 0.1--100~keV.
Thus,  we need the value of $S(0)$ in practical calculations
as $S(E)$ is a very slowly varying function of energy at low energy region.%
\footnote{The cold and hot CNO burning cycles have the Gamow peak energy regions at $E_G \simeq 26$~keV and $150-200~\mathrm{keV}$ 
for the corresponding core temperatures at $T_c \simeq 1.5\times 10^7$~K and $\mbox{(2-3)} \times 10^8$~K, respectively~\cite{WGUIP10,JH97}. }
Therefore, most realistic method for obtaining the astrophysical $S$ factor at zero energy would be the extrapolation of experimentally 
determined $S(E)$ to the lower energy range.
However, because of the difficulties of experiments near threshold, larger error bars at low energy region are inevitable and this causes 
the uncertainties of $S(0)$.
The extrapolation would, of course, depend on the theoretical models adopted for the description of the reaction.

Despite the importance of the $\nuclide[15]{N} (p, \gamma) \nuclide[16]{O}$ reaction, experimental measurements of its cross sections at low
energies are rare and only two measurements were reported until 2009~\cite{Hebbard60,RR74}.%
\footnote{The experimental data reported in Ref.~\cite{SFL52} are for $E_p \ge 860$~keV and do not constrain the value of $S(0)$,
where $E_p$ is the kinetic energy of the incident proton in the laboratory frame.
We also note that the data of Ref.~\cite{BCDRS73} covered the region of $E_p > 250$~keV.}
The first measurement was made about 60 years ago~\cite{Hebbard60}, which was then followed by the work of Ref.~\cite{RR74} in mid 1970s.
The former experiment reported $S(0) \simeq 32$~$\mbox{keV} \  \mbox{b}$, while the latter estimated 
$S(0) = 64 \pm 6$~$\mbox{keV} \  \mbox{b}$.
This shows that the value of the latter experiment is about twice that of the former one.
However, these experiments verified that the reaction is dominated by the first two interfering $J^\pi = 1^-$ resonances at $E_R = 312$ and $964$~keV, 
respectively, where $E_R$ is the resonance energy in the center-of-mass frame~\cite{MBBG08}.%
\footnote{The corresponding excitation energies are 12.45~MeV and 13.09~MeV, respectively.
In the laboratory frame, these values correspond to $E_p = 338$~keV and $1028$~keV, respectively.}
These two data sets are used for the analyses in Refs.~\cite{Barker08b,MBBG08,HBG08}.
In particular, it was pointed out in Ref.~\cite{MBBG08} that the contribution from the direct capture process was overestimated in Ref.~\cite{RR74}.

In the last decade, more measurements were performed at the Laboratory for Underground Nuclear Astrophysics (LUNA) at 
Gran Sasso underground laboratory (LNGS) and at the Notre Dame Nuclear Science Laboratory~\cite{BCBB09,LIGJ10,CMCB11}.
These efforts are summarized in Refs.~\cite{LUNA11,LUNA17} and the data are collected, for example, in Refs.~\cite{XTGAOU13,EXFOR} which
are used for updating the estimation of the $S$ factors in Refs.~\cite{MLK11,IDBC12,dGILUW13,DD14}.%
\footnote{The experimental data of Ref.~\cite{LIGJ10} are compiled in Ref.~\cite{dGILUW13}. 
In the present work, we use the experimental data compiled in Experimental Nuclear Reaction Data (EXFOR)~\cite{EXFOR}.}
The newly estimated values of $S(0)$ based on these data sets, called the post-NACRE data, are as low as $33.1$~$\mbox{keV} \  \mbox{b}$~\cite{MLK11} 
and as high as $45^{+9}_{-7}$~$\mbox{keV} \  \mbox{b}$~\cite{XTGAOU13}.

\begin{figure}[t]
\centering
\includegraphics[width=0.5\textwidth]{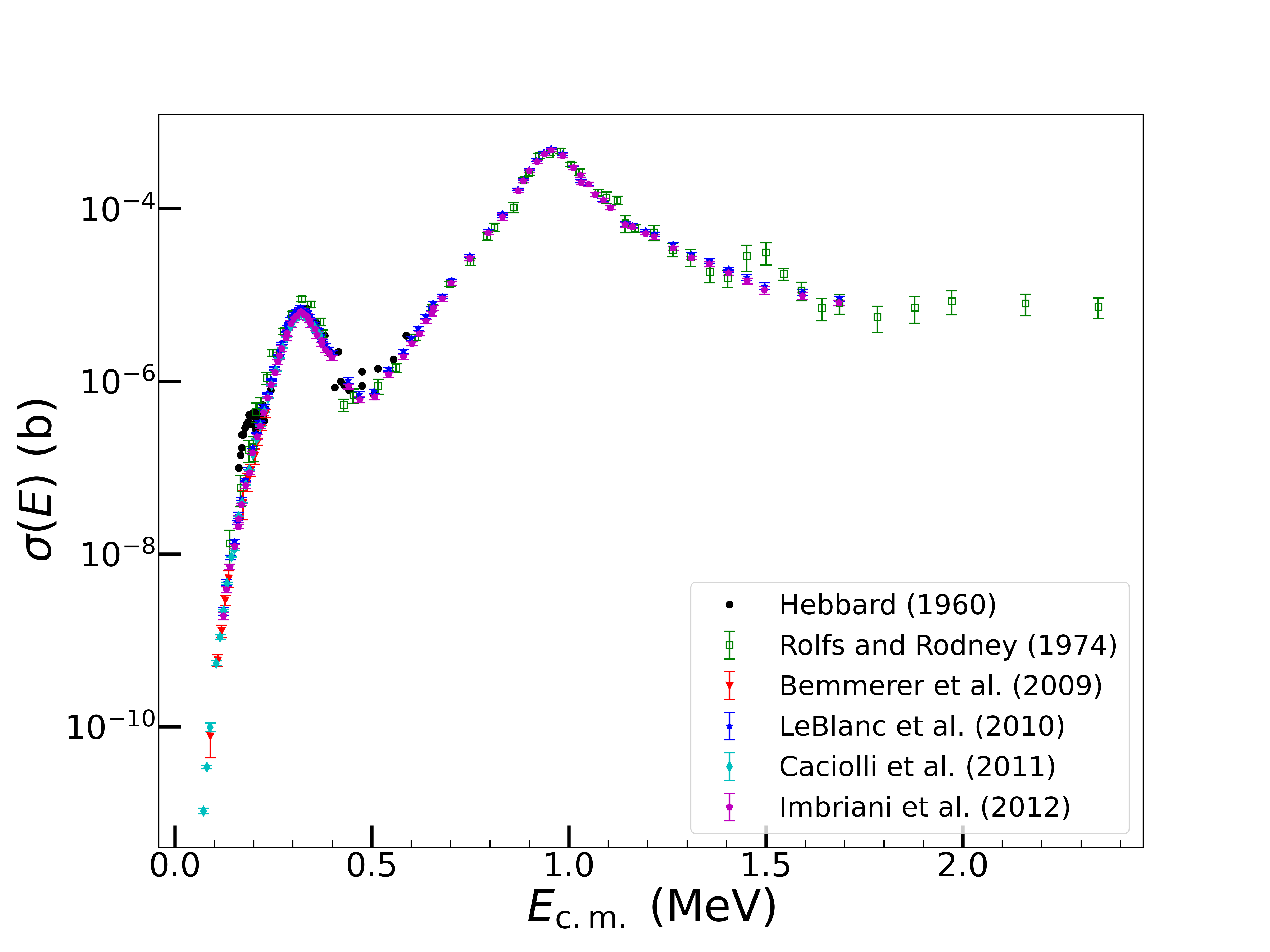}
\includegraphics[width=0.5\textwidth]{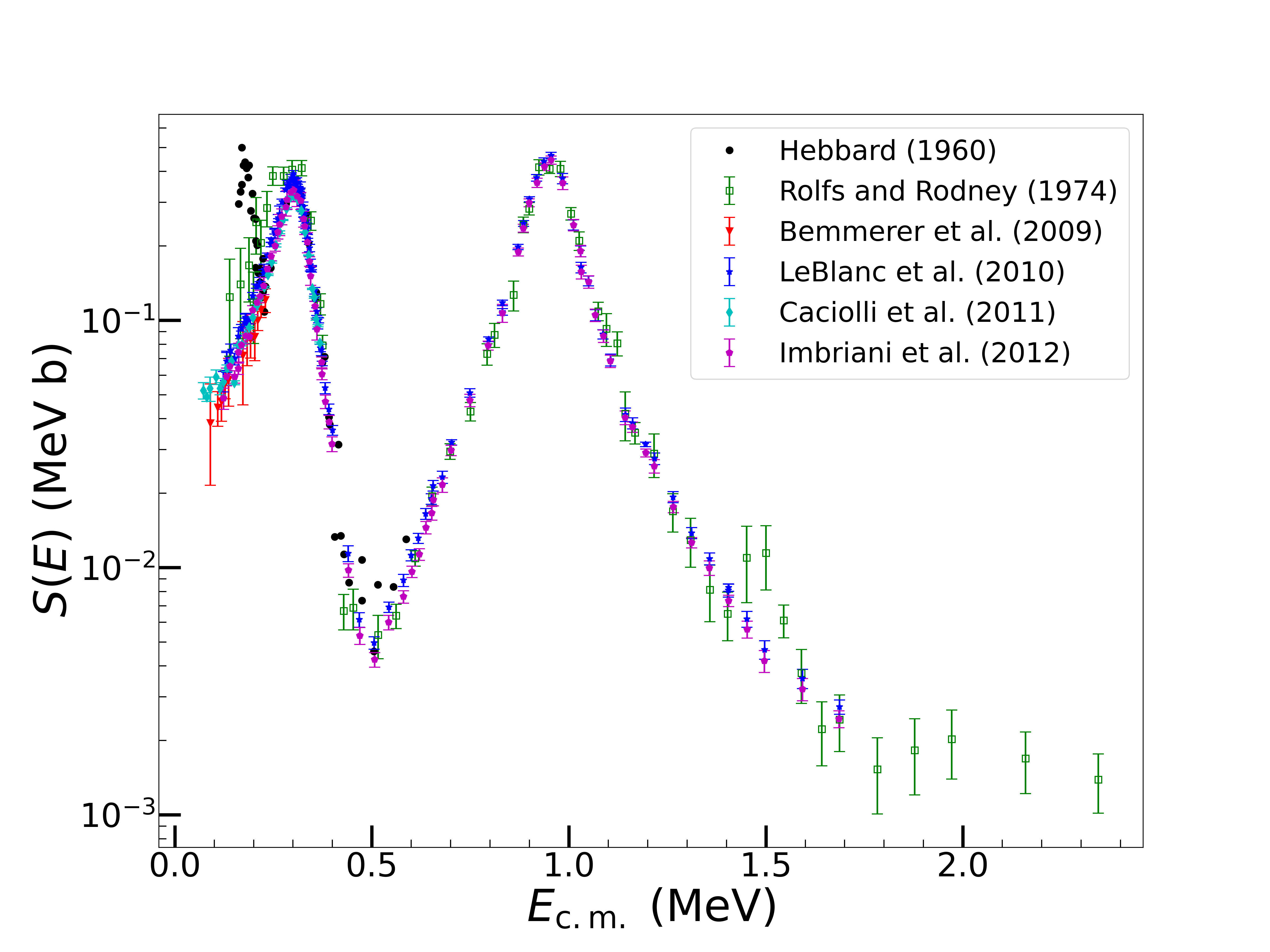}
\caption{Collected experimental data for (a) the cross sections and (b) the astrophysical $S$ factors $S(E)$ of 
the $\nuclide[15]{N} (p, \gamma) \nuclide[16]{O}$ reaction.
The data from Refs.~\cite{Hebbard60,RR74,BCBB09,LIGJ10,CMCB11,IDBC12} are compiled in Ref.~\cite{EXFOR}.
}
\label{fig:data}
\end{figure}

Shown in Fig.~\ref{fig:data} are currently available experimental data of $\nuclide[15]{N} (p, \gamma) \nuclide[16]{O}$
compiled in Ref.~\cite{EXFOR}.
Figure~\ref{fig:data}(a) shows the total cross section data, while Fig.~\ref{fig:data}(b) presents the derived astrophysical $S$ factors.
They are given as functions of $E_{\rm cm}$, the kinetic energy of the system in the center-of-mass frame.
These figures explicitly show that the reaction is dominated by the two broad resonances mentioned previously.
Except the earlier data of Refs.~\cite{Hebbard60,RR74}, the reported data show a good agreement.
The cross section data of Ref.~\cite{Hebbard60} indicate a small structure at vey low energies and it is further exaggerated in the
astrophysical $S$ factor as shown by black dots in Fig.~\ref{fig:data}(b).
It is also evident that the data of Ref.~\cite{RR74} have larger error bars compared with the data of the next generation experiments.
Furthermore, the data from Ref.~\cite{CMCB11, BCBB09} do not cover the second resonance region.
Therefore, in the present work, we focus on the data covering the both resonance regions, which include the data of Ref.~\cite{RR74} 
and the post-NACRE data of Refs.~\cite{LIGJ10,IDBC12}, and apply our calculations to these three experimental data sets.

In the present work, we apply the effective field theory (EFT) formalism to investigate the \nuclide[15]{N}($p$,$\gamma$)\nuclide[16]{O} reaction.
EFTs allow for a systematic calculation by introducing a scale $\Lambda_H$ which separates relevant degrees of freedom 
at low energies from irrelevant degrees of freedom at high energies.
Then the effective Lagrangian is constructed by the expansion with the number of derivatives, and the reaction amplitudes 
are calculated in powers of $Q/\Lambda_H$ where $Q$ is a typical momentum scale of the reaction in question. 
The irrelevant high energy degrees of freedom are integrated out and the coefficients in the effective Lagrangian, called 
low-energy constants (LECs), are determined by fitting the experimental data~\cite{Weinberg79}.
EFT methods have been applied to the studies of various reactions including thermonuclear reactions such as $\alpha$ capture 
on \nuclide[12]{C}~\cite{Ando18}, 
radiative proton capture on \nuclide[12]{C}~\cite{KKS17}, and proton and neutron capture on light 
nuclei~\cite{ZNP13,ZNP15,ZNP17,RH11,HRV16}.
It was also applied to the reactions of \nuclide[15]{C}~\cite{MYC19}, $\alpha$-\nuclide[12]{C} elastic
scattering~\cite{Ando16,Ando18b,Ando21}, and $\beta$ delayed $\alpha$ emission from \nuclide[16]{N}~\cite{Ando20b}, and so on.
Reviews on these topics can be found, for example, in Refs.~\cite{BV02c,Rupak16,HJP17,Capel21}.

The construction of EFT for the $\nuclide[15]{N} (p, \gamma) \nuclide[16]{O}$ reaction involves three open channels, namely, $p$-\nuclide[15]{N}, 
$\alpha$-\nuclide[12]{C}, and $\alpha$-$\nuclide[12]{C}^*$ where $\nuclide[12]{C}^*$ denotes the first excited $2^+$ state of \nuclide[12]{C}. 
In the present study, by focusing on the $\nuclide[15]{N} (p, \gamma) \nuclide[16]{O}$ reaction data, we perform the single channel
calculations leaving the coupled-channel calculations to a future work.
The energy range of the data of our interests covers up to $E \simeq 2$~MeV,
which includes the aforementioned two $s$-wave resonant states of \nuclide[16]{O}. 
The inclusion of a resonant state in EFT was investigated, for example, by Gelman~\cite{Gelman09} and by Habashi, Fleming, 
and van Kolck~\cite{HFV20}, and, following their method, we here employ the effective Lagrangian for the $p$-\nuclide[15]{N} system 
with the two resonant states of \nuclide[16]{O}.

We choose the breakup energy of \nuclide[16]{O} into the $n$-\nuclide[15]{O} channel, $\Delta E=3.54$~MeV, as the large energy scale,
which gives $\Lambda_H=\sqrt{2\mu \Delta E}=80$~MeV, where $\mu$ is the reduced mass of the system. 
On the other hand, the resonant energies, $E_1 = 0.312$~MeV and $E_2=0.964$~MeV, are chosen as the typical energy scales of the theory
leading to $Q_1 = \sqrt{2\mu E_1}= 24$~MeV and $Q_2 = \sqrt{2\mu E_2}=41$~MeV.
Therefore, our expansion parameters are $Q_1/\Lambda_H \simeq 0.3$ and $Q_2/\Lambda_H\simeq 0.5$, and
the terms in the amplitudes are expanded in powers of $(Q_{1,2}/\Lambda)^{2n}$ as the effective range expansion (ERE). 
We include the terms up to the next-to-leading order for the first resonant state and up to the fourth order terms for the second so that
the theoretical uncertainties of the present calculations are estimated as $(Q_1/\Lambda_H)^2 \simeq 0.09$ and $(Q_2/\Lambda_H)^6 \simeq 0.016$.

Our preliminary results in this approach were reported in Ref.~\cite{SAO22a} where only the first resonance was considered with
the experimental data of Ref.~\cite{CMCB11}.
The estimated astrophysical $S$ factor is $S(0) = 30.4$~$\mbox{keV} \  \mbox{b}$ that is close to the lower limit of $S(0)$
obtained in the recent analyses of Ref.~\cite{MLK11}.
In the present work, we perform more complete analysis working with the data covering the energy regions of both resonances which
were reported in Refs.~\cite{RR74,LIGJ10,IDBC12}.

This paper is organized as follows.
In the next section, we develop the cluster EFT formalism for the reaction of $\nuclide[15]{N} (p, \gamma) \nuclide[16]{O}$
by introducing the effective Lagrangian for this reaction.
In Sec.~\ref{sec:di-field}, the propagator of the di-field, which is introduced for the description of resonances, is discussed.
Then the capture amplitudes are derived in Sec.~\ref{sec:amplitude}, and
Sec.~\ref{sec:results} presents our numerical results which are compared with the estimates in the literature.
Section~\ref{sec:summary} summarizes this work and the derivation of the loop integral formulas is given in Appendix.

\section{Effective Lagrangian}
\label{sec:Lag}

The cluster EFT, which is applied to the investigation of nuclear reactions at stellar environments, is similar to the pionless 
EFT~\cite{BHV02, BHV03} by adopting contact couplings among the participating particles. 
In the present approach, we treat the nuclei involved in the the reaction of $\nuclide[15]{N}(p, \gamma)\nuclide[16]{O}$ 
as point-like particles, and describe resonance states $\nuclide[16]{O}^*$ as bound systems of the proton and the 
$\nuclide[15]{N}$ nucleus.
The high energy scale is determined by the breakup energy of \nuclide[16]{O} and the resonant energies are used to 
estimate the low energy scales.%
\footnote{The momentum scale corresponding to the radius of $\nuclide[16]{O}$ is about 70 MeV which is larger than our typical 
scales and is comparable to the large scale.
The finite-range effects would be investigated by the expansion in powers of spatial derivatives.}

 Since we have two resonant $\nuclide[16]{O}^*$, we construct the effective Lagrangian for this reaction as
\begin{widetext}
\begin{eqnarray} \label{eq:Lagrangian}
\mathscr{L} &=& \psi_p^\dagger \left\{ iv\cdot D + \frac{1}{2 M_p} \left[ (v\cdot D)^2-D^2 \right] \right\} \psi_p 
+ \psi_N^\dagger \left\{ iv \cdot D + \frac{1}{2 M_N} \left[ (v\cdot D)^2-D^2 \right] \right\} \psi_N 
\nonumber \\ && \mbox{} 
+ \sum_{n=0}^{n_{\rm max}} C_n^{(I=0)} d_i^{(I=0) \dagger} \left\{ iv\cdot D + \frac{1}{2(M_p + M_N)} \left[ (v\cdot D)^2-D^2 \right] \right\}^n d_i^{(I=0)} 
\nonumber \\ && \mbox{} 
+ \sum_{m=0}^{m_{\rm max}} C_m^{(I=1)} d_{ia}^{(I=1) \dagger} \left\{ iv\cdot D + \frac{1}{2(M_p + M_N)} \left[ (v\cdot D)^2-D^2 \right] \right\}^m d_{ia}^{(I=1)} \nonumber \\
&& \mbox{} - y_t^{(I=0)} \biggl\{ d_i^{(I=0)\dagger} \biggl( \psi_p^T P^{(^{3}S_1)}_i \psi_N \biggr) + \mbox{H.c.} \biggr\} 
- y_t^{(I=1)} \biggl\{ d_{ia}^{(I=1) \dagger} \biggl( \psi_p^T P^{(^{3}S_1)}_{ia}\psi_N\biggr)+\mbox{H.c.}\biggr\} \nonumber \\
&& \mbox{} -y_s \biggl\{\phi_O^\dagger\biggl(\psi_p^TP^{(^{3}P_0)}\psi_N \biggr)+\mbox{H.c.}\biggr\} 
- y_t^{(I=0)} h^{(I=0)} \biggl\{ \phi_O^\dagger (-iD_i) d_i^{(I=0)} + \mbox{H.c.} \biggr\} \nonumber \\
&& \mbox{} - y_t^{(I=1)} h^{(I=1)} \biggl\{ \phi_O^\dagger(-iD_i) d_{ia}^{(I=1)} \delta_{a3} + \mbox{H.c.} \biggr\},    
\end{eqnarray}
\end{widetext}
where $\psi_p$, $\psi_N$, and $\phi_O$ are the fields of the proton, $\nuclide[15]{N}$, and the ground state of $\nuclide[16]{O}$, respectively,
and the masses of the proton and \nuclide[15]{N} are denoted by $M_p$ and $M_N$, respectively.
The interactions with higher derivatives are suppressed in the effective Lagrangian of Eq.~(\ref{eq:Lagrangian}).
The di-fields $d^{(I=0)}$ and $d^{(I=1)}$ are introduced as auxiliary fields to describe the first and second resonance states of 
$\nuclide[16]{O}$, respectively, where $I$ in the superscript indicates its isospin quantum number.
The coupling constants $y_s$, $y_t^{(I)}$, and $h^{(I)}$ are the LECs which determine the strength of the 
contact interactions among the proton, $\nuclide[15]{N}$, and $\nuclide[16]{O}$ nucleus.

\begin{figure*}[t]
\centering
\includegraphics[width=0.7\textwidth]{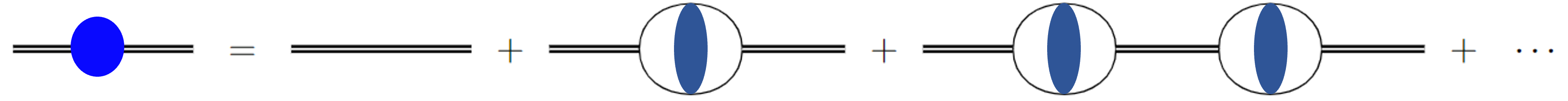}
\caption{Fully-dressed propagator of the di-field. 
In the right hand side, the double line stands for the bare propagator and the shaded region represents the Coulomb interaction 
between two nuclei.}
\label{fig:propagator}
\end{figure*}

The four-velocity vector $v^\mu$ is chosen to be $v^\mu = (1, \bm{0})$, and $D_\mu = \partial_\mu + ie\hat{Q}A_\mu$ 
is the covariant derivative where $e$, $\hat{Q}$, and $A_\mu$ are the elementary charge, charge operator, 
and the photon field, respectively. 
The projection operators in the interaction terms are defined as
\begin{eqnarray}
P^{(^3S_1)}_i &=& \frac{1}{2} \sigma_2^{} \sigma_i^{} \tau_2^{}, \qquad 
P^{(^3S_1)}_{ia} = \frac{1}{2} \sigma_2^{} \sigma_i^{} \tau_2^{} \tau_a^{}, \\
P^{(^3P_0)} &=&  \frac{1}{2} \left( \tau_2^{} \sigma_2^{} \vec{\sigma}^{} \cdot \frac{(-i\overrightarrow{D})}{M_N} 
- \frac{(-i\overleftarrow{D})}{M_p} \cdot \tau_2^{} \sigma_2^{} \vec{\sigma} \right),
\end{eqnarray}
where the superscript in the projection operator stands for the orbital angular momentum of the interaction between 
the proton and the $\nuclide[15]{N}$ nucleus. 
Here, $\sigma_i^{}$ are the Pauli spin matrices and $\tau_i^{}$ are the Pauli isospin matrices.
The integers $n_{\rm max}$ and $m_{\rm max}$ are the numbers of the considered effective range parameters of 
the effective range expansion (ERE) which determine the coefficients $C^{(I=0)}_n$ and $C^{(I=1)}_m$ for 
the iso-singlet and iso-triplet di-fields, respectively.
For the two resonances these integers are chosen by $n_{\rm max} = 1$ and $m_{\rm max} = 3$ 
corresponding to the terms up to the second and the fourth order.%
\footnote{As we mentioned, the orders associated with the two resonances differ. 
In order to further exploit the uncertainty control, we have increased the order with the lower resonance to $(Q_1/\Lambda_H)^4$
so that the theoretical uncertainties from the two resonances are similar, namely, 0.008 and 0.016, respectively. 
However, we could not find any improvement with this choice and we have a feature of over-fitting. 
Through these trials, we found that ($n_{\rm max} = 1$, $m_{\rm max} = 3$) is the optimal choice.}

\section{Di-field propagator}
\label{sec:di-field}

In the present work, we depict the intermediate resonance states by considering the fully-dressed propagator of di-field $d^{(I)}$. 
As shown in Fig.~\ref{fig:propagator}, the fully-dressed propagator is expressed as an infinite series of the bare propagator $D_0$ 
and the self-energy $\Sigma$ as
\begin{equation}
    D = \left( D_0^{-1}-\Sigma \right)^{-1}.
\end{equation}
In this figure, the self-energy is represented by the bubble diagram with a shaded region between the two bare propagators. 
From the Lagrangian (\ref{eq:Lagrangian}), the bare propagator for $d^{(I)}$ is written as
\begin{equation}
    D_0^{(I)}(p) = \frac{1}{C_0^{(I)} + C_1^{(I)}p^2 + C_2^{(I)}p^4 + \cdots},
\end{equation}
where $p$ is the magnitude of the relative spatial momentum between the two nuclei in the c.m. frame. 
Following the approach of Refs.~\cite{KR98b, KR99}, the Coulomb interactions between the two nuclei at low-energies are 
described by using the Coulomb Green's function $G_C(E)$, which leads to
\begin{eqnarray}
    \Sigma_{ij}(E) &=& \frac{1}{2} \left( y_t^{(I)} \right)^2  \Braket{ 0 | G_C(E) | 0} \delta_{ij}  \nonumber \\
    &=& \frac{1}{2} \left( y_t^{(I)} \right)^2 \left(-\frac{\mu}{\pi}\kappa H(\eta) + J_0^{\rm div}\right) \delta_{ij} ,
\end{eqnarray}
where $E = p^2/2\mu$ is the total kinetic energy and
\begin{equation}
    H(\eta) = \psi(i\eta) - \ln(i\eta) - \frac{i}{2\eta},
\end{equation}
where $\psi(z)$ is the digamma function defined as $\psi(z) = d [ \ln\Gamma(z) ] / dz$ with $\Gamma(z)$ being the gamma function.
The Sommerfeld parameter $\eta$ is defined as $\eta = \kappa/p$ with $\kappa = Z_pZ_N\mu\alpha_{\rm em}$, 
where $Z_A$ is the charge number of nucleus $A$. 
In our case, $\kappa = 44.9$~MeV.

The divergent part $J_0^{\rm div}$ arising from the Coulomb Green's function reads
\begin{equation}
    J_0^{\rm div} = \frac{\mu}{\pi}\kappa\left(\frac{1}{\epsilon} + \ln\frac{\Lambda\sqrt{\pi}}{2\kappa} + 1 -\frac{3}{2}\gamma_E^{} 
    - \frac{\Lambda}{2\kappa}\right),
    \label{J0div}
\end{equation}
where $\gamma_E^{}$ is the Euler's constant and $\Lambda$ is the regularization scale introduced by performing the dimensional regularization 
in space-time dimension of $d = 4-\epsilon$. 
The term linear in $\Lambda$ comes from the power divergence subtraction scheme~\cite{KSW98a}. 
One may write the fully-dressed propagator by making use of the ERE so that the coefficients $C^{(I)}_i$ can be fixed 
by the effective range parameters~\cite{AH04}, which gives
\begin{equation}
    D^{(I)}(p) = - \left( y_t^{(I)} \right)^{-2} \frac{4\pi}{\mu}\frac{1}{K^{(I)}(p)-2\kappa H(\eta)},
    \label{eq:prop}
\end{equation}
and the ERE allows to write
\begin{equation}
    K^{(I)}(p) = -\frac{1}{a_R^{(I)}} + \frac{1}{2}r^{(I)}p^2 - \frac{1}{4}P^{(I)}p^4 + \cdots,
\end{equation}
where $a_R^{(I)}$, $r^{(I)}$, and $P^{(I)}$ are the scattering length, effective range, and the shape parameter, respectively, 
and the divergent part $J_0^{\rm div}$ of Eq.~(\ref{J0div}) is absorbed by the scattering length.

Following Ref.~\cite{HHV08}, the effective range parameters appearing in the intermediate propagator 
are related to the resonance energy and its decay width. 
It is achieved by comparing the obtained amplitude of the elastic scattering process with the one obtained with the Breit-Wigner form. 
In Refs.~\cite{Ando20a, Ando21}, it is modified by including additional terms in the denominator of the propagator 
as a result of expanding the denominator near the resonance energy. 
The real part of the denominator of the propagator is then expanded as
\begin{eqnarray}
    \Tilde{K}^{(I)} &=& K^{(I)} - 2\kappa \, \mathrm{Re}\,H(\eta) \nonumber \\
    &=& -\frac{1}{a_R^{(I)}} + \frac{1}{2}\Tilde{r}^{(I)}p^2 -\frac{1}{4}\Tilde{P}^{(I)}p^4 + \Tilde{Q}^{(I)}p^6 
\label{eq:ERE}
\end{eqnarray}
up to $O(p^6)$, where%
\footnote{This is an expansion by $p/\kappa$ and, since $p$ of the second resonance region is close to the value of $\kappa$, the expansion
would be questionable. In our case, however, the expansion of the function $\mbox{Re}\, H(\eta)$ converges fast and is valid within 
5\% even in the second resonance region ($E_{\rm cm} \lesssim 1.5$~MeV).}
\begin{eqnarray}
\Tilde{r}^{(I)} &=& r^{(I)}-1/3\kappa, 
\nonumber \\
\Tilde{P}^{(I)} &=& P^{(I)} + 1/15\kappa^3,
\nonumber \\ 
\Tilde{Q}^{(I)} &=& Q^{(I)} - 1/126\kappa^5.
\end{eqnarray}
Here, the term containing $\kappa$ comes from the Coulomb self-energy, $-2\kappa H(\eta)$.
The imaginary part of the propagator is $-2\kappa \,\mbox{Im} [ H(\eta)] = -p C_\eta^2$, where
\begin{equation}
C_\eta = \sqrt{\frac{2\pi\eta}{e^{2\pi\eta}-1}}
\label{eq:Gamow}
\end{equation} 
is the Gamow factor.

\begin{figure*}[t]
\centering
\includegraphics[page=1,width=\textwidth]{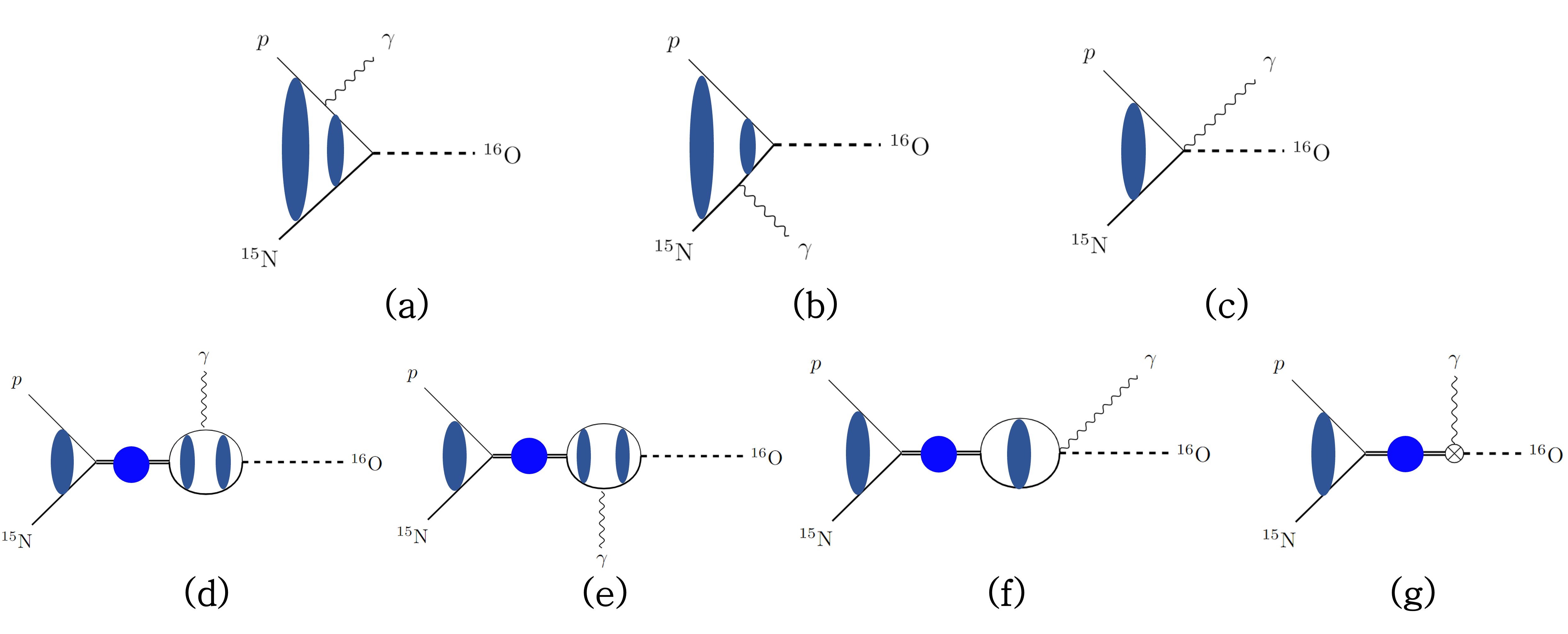}
\caption{Feynman diagrams for the radiative capture process of $\nuclide[15]{N}(p, \gamma)\nuclide[16]{O}$. 
The thin and thick solid lines denote the proton and the \nuclide[15]{N} nucleus, respectively, and the wavy line stands for the out-going photon. 
The dotted line represents the ground state of the \nuclide[16]{O} nucleus in the final state. 
The shaded blobs represent the Coulomb repulsion between the proton and \nuclide[15]{N}. 
The circle with a cross in (g) is introduced as a counterterm of the loop diagrams.}
\label{fig:npgo}
\end{figure*}

Considering up to the order of $p^6$ in the ERE, the real part of the denominator of propagators can be expanded around the resonance energy as
\begin{equation}
    \Tilde{K}^{(I)}(E) = \left. \sum_{n=0}^3 \frac{1}{n!}\frac{\partial^n \Tilde{K}^{(I)}}{\partial E^n} \right |_{E=E_R} (E-E_R)^n,
\end{equation}
with $\Tilde{K}^{(I)}(E_R) = 0$ where $E_R$ is the resonance energy with $p_R^{} = \sqrt{2 \mu E_R}$.
The di-field propagator with the effective range parameters can also be represented in terms of the resonance energy and width as the 
Breit-Wigner formula, which reads
\begin{equation}
    D(p) = \left(y_t^{(I)}\right)^{-2} \frac{4\pi}{\mu p} \frac{\frac12\Gamma_R(E) C_\eta^{-2}}{E - E_R + R(E) + i \frac12 \Gamma_R(E)}, 
    \label{propagator}
\end{equation}
with $\eta_R^{} = \kappa/p_R^{}$, and the energy dependence of the resonance width is written as
\begin{eqnarray}
    \Gamma_R(E) &=& -\frac{4\pi\kappa}{\mu \left( \Tilde{r} - \Tilde{P} p_R^2 + 6\Tilde{Q}p_R^4 \right) }  \frac{1}{e^{2\pi\eta}-1}
    \nonumber \\
    &=& \Gamma_R(E_R)\frac{e^{2\pi\eta_R^{}}-1}{e^{2\pi\eta}-1} . \label{width}
\end{eqnarray}
In Eq.~(\ref{propagator}), $R(E)$ contains the higher order corrections in the expansion around the resonance energy, which reads 
\begin{equation}
    R(E) = a(E-E_R)^2 + b(E-E_R)^3,
\end{equation}
where 
\begin{eqnarray}
    a = \frac{-\mu \Tilde{P} + 12\mu \Tilde{Q}p_R^2}{\Tilde{r} - \Tilde{P} p_R^2 + 6\Tilde{Q}p_R^4}, \quad 
    b = \frac{8\mu^2 \Tilde{Q}}{\Tilde{r} - \Tilde{P} p_R^2 + 6\Tilde{Q}p_R^4}. \label{eq:ab}
\end{eqnarray} 
Through Eqs.~(\ref{propagator})--(\ref{eq:ab}) we omit isospin index $I$ for simplicity. 
We take the resonance energy, width, $a$, and $b$ as free parameters to be fixed by fitting the experimental data of the astrophysical $S$ factors,
which will then give the effective range parameters through Eqs.~(\ref{eq:ERE}), (\ref{width}), (\ref{eq:ab}), 
and the condition that $\Tilde{K}(E_R)=0$.

\section{Capture amplitudes}
\label{sec:amplitude}

The diagrams of the radiative capture amplitudes for the $\nuclide[15]{N}(p, \gamma)\nuclide[16]{O}$ reaction are 
depicted in Fig.~\ref{fig:npgo}. 
The capture amplitude $\mathscr{M}$ can be decomposed into the iso-singlet and iso-triplet parts as
\begin{equation}
    \mathscr{M} = \frac{e}{2}\chi_p^T \sigma_2^{} (\bm{\epsilon}_\gamma^*\cdot\bm{\sigma}) \tau_2^{} 
    \left(X^{(I=0)} - \tau_3^{} X^{(I=1)} \right) \chi_N^{}, 
    \label{eq:amplitude}
\end{equation}
where the spinors for the proton and the $\nuclide[15]{N}$ nucleus in the initial state are represented by $\chi_p^{}$ and $\chi_N^{}$, respectively, 
and $\bm{\epsilon}_\gamma$ is the photon polarization vector.
In Fig.~\ref{fig:npgo}, the shaded blobs represent the Coulomb repulsion between the proton and \nuclide[15]{N}. 
The circle with a cross ($\otimes$) in Fig.~\ref{fig:npgo}(g) is the photon coupling between \nuclide[16]{O} and the di-fields which is introduced as a counterterm of the loop diagrams.
The isospin-dependent terms can be written as
\begin{eqnarray}
    X^{(I=0)} &=& X_{(a+b)} + X_{(c)} + X^{(I=0)}_{(d+e)} + X^{(I=0)}_{(f)} + X^{(I=0)}_{(g)}\,, \label{X0} \\
    X^{(I=1)} &=& X^{(I=1)}_{(d+e)} + X^{(I=1)}_{(f)} + X^{(I=1)}_{(g)}\, \label{X1},
\end{eqnarray}
where the amplitudes, $X_{(a,b,c)}$, $X^{(I=0)}_{(d,e,f)}$, and $X^{(I=1)}_{(d,e,f)}$ are calculated from the corresponding Feynman diagrams
in Fig.~\ref{fig:npgo}.

In order to obtain the radiative capture amplitudes, we follow the approach advocated by the authors of Refs.~\cite{RFHP13, RFHP14}. 
Using the effective Lagrangian of Eq.~(\ref{eq:Lagrangian}), the amplitude for non-resonant process depicted in Fig.~\ref{fig:npgo}(a) 
in the c.m. frame is obtained as
\begin{eqnarray}
\mathscr{M}_{(a)} &=& -\frac{y_s}{\mu}\frac{Z_p}{M_p} I_{(a)}^{ij}\frac{e}{2} \chi_p^T \epsilon_\gamma^{*i} \sigma_2^{} \sigma_j^{} \tau_2^{} \chi_N^{},
\label{A}
\end{eqnarray}
where $I_{(a)}^{ij}$ comes from the loop integration defined as
\begin{eqnarray}
    I_{(a)}^{ij} &=& \int\frac{d^3\ell \, d^3 \ell'}{(2\pi)^6} d^3r \, d^3r' \, d^3r'' \, \ell_i \, \ell'_j \Braket{ \bm{ \ell' } | {\bf r}' }  
    \nonumber \\ && \mbox{} \times
    \Braket{ {\bf r}' | G_C(-B) | {\bf r} }
    \Braket{ {\bf r} | \bm{\ell} - \frac{\mu}{M_p} {\bf k} }
    \Braket{ \bm{\ell} | {\bf r}'' } \Braket{ {\bf r}'' | \psi_{\bf p} }.
    \nonumber \\ 
\end{eqnarray}
Here, $B$ is the binding energy of $\nuclide[16]{O}$ relative to the $p$-$\nuclide[15]{N}$ breakup threshold, 
${\bf k}$ is the outgoing photon momentum, and $\Braket{ {\bf r} | \psi_{\bf p} } = \psi_{\bf p} ({\bf r})$ is the Coulomb wave function. 
One can expand the Coulomb wave function and Green's function by partial waves as
\begin{eqnarray}
    \psi_{\bf p} ({\bf r}) &=& \sum_{l=0}^{\infty} (2l+1) i^l e^{i\sigma_l} P_l(\hat{\bf r} \cdot \hat{\bf p}) \frac{F_l(\eta, \rho)}{\rho}, \\
    \Braket{ {\bf r}' | G_C(E) | {\bf r} } &=& \sum_{l=0}^{\infty} (2l+1) G_C^{(l)}(E; r', r) P_l(\hat{\bf r}' \cdot \hat{\bf r}), 
    \label{G_partial}
\end{eqnarray}
where $\rho = pr$, $\sigma_l = \mbox{arg}\left[ \Gamma(l+1+i\eta) \right]$ is the Coulomb phase shift, 
$P_l(x)$ is the Legendre polynomial, and $F_l(\eta, \rho)$ is the regular Coulomb function. 
It can be easily verified that the $s$-wave ($l=0$) contribution in Eq.~(\ref{G_partial}) vanishes by symmetry consideration
and only the $p$-wave ($l=1$) part contributes. 
In Ref.~\cite{RFHP14}, the partial-wave-expanded Coulomb Green's function $G_C^{(l)}$ for the bound state is shown to be 
simplified when it is written in terms of the regular and irregular Coulomb functions, $F_l(\eta, \rho)$ and $G_l(\eta, \rho)$, which leads to
\begin{equation}
    G_C^{(l)}(-B, \rho', \rho) = -\frac{\mu p}{2\pi}\frac{F_l(\eta, \rho')}{\rho'}\frac{iF_l(\eta, \rho)+G_l(\eta, \rho)}{\rho}.
\end{equation}
Using the identities of the Coulomb functions~\cite{ORBC} the loop integration $I_{(a)}^{ij}$ is rewritten as
\begin{eqnarray}
    I_{(a)}^{ij} &=& \delta_{ij} \frac{2\mu\gamma}{3} \Gamma(2+\kappa/\gamma) e^{i\sigma_0}
    \nonumber \\ && \mbox{} \times 
    \int_0^\infty dr rW_{-\kappa/\gamma, 3/2}(2\gamma r)
    j_0\left(\frac{\mu}{M_p}kr\right)\frac{\partial}{\partial r}\left(\frac{F_0(\eta, pr)}{pr}\right), 
    \nonumber \\
\end{eqnarray}
where $j_l(z)$ and $W_{i\eta, l+1/2}(2i\rho)$ are the spherical Bessel function and the Whittaker $W$-function, respectively. 
Here, $\gamma = \sqrt{2\mu B}$ is the binding momentum determined by the binding energy $B$ of \nuclide[16]{O} relative to the $p$-\nuclide[15]{N} threshold. 
Hence, the amplitude $\mathscr{M}_{(a)}$ is obtained as
\begin{eqnarray}
    \mathscr{M}_{(a)} &=& -y_s \frac{2\gamma Z_p}{3 M_p} \Gamma(2+\kappa/\gamma) e^{i\sigma_0}
        \nonumber \\ && \mbox{} \times  
    \int_0^\infty dr r W_{-\kappa/\gamma, 3/2}(2\gamma r) 
    j_0\left(\frac{\mu}{M_p}kr\right)\frac{\partial}{\partial r} \left(\frac{F_0(\eta, pr)}{pr}\right) 
    \nonumber \\ && \mbox{} \qquad \times 
    \chi_p^T \frac{e}{2} \sigma_2^{} \left( \bm{\epsilon}_\gamma^* \cdot \bm{\sigma} \right) \tau_2^{} \chi_N^{}.
\end{eqnarray}
The rest of the amplitudes can be calculated in a similar way, and the isospin-dependent terms in Eq.~(\ref{eq:amplitude}) are obtained as 
\begin{widetext}
\begin{eqnarray}
X_{(a+b)} &=& y_s \frac{2\gamma}{3} \Gamma(2+\kappa/\gamma) e^{i\sigma_0} 
\nonumber \\ && \mbox{} \times 
\int_0^{\infty} dr r W_{-\kappa/\gamma, 3/2}(2\gamma r)
\left[ \frac{Z_N}{M_N} j_0\left(\frac{\mu}{M_N}kr\right) - \frac{Z_p}{M_p} j_0\left(\frac{\mu}{M_p}kr\right) \right]
\frac{\partial}{\partial r} \left(\frac{F_0(\eta, pr)}{pr}\right), 
\label{eq:X_ab} \\
X_{(c)} &=& y_s \left( \frac{Z_N}{M_N} - \frac{Z_p}{M_p} \right) C_\eta e^{i\sigma_0}, 
\\
X^{(I)}_{(d+e)} &=& y_s \frac{2\gamma}{3} \Gamma(1+i\eta) \Gamma(2+\kappa/\gamma) 
\frac{C_\eta e^{i\sigma_0}}{K^{(I)}(p)-2\kappa H(\eta)} 
\nonumber \\ &&\mbox{} \times 
\int_{0}^\infty dr r W_{-\kappa/\gamma, 3/2}(2\gamma r) \left[ \frac{Z_N}{M_N} j_0 \left(\frac{\mu}{M_N} kr\right) - \frac{Z_p}{M_p} 
j_0\left(\frac{\mu}{M_p} kr\right) \right]
\frac{\partial}{\partial r} \left(\frac{W_{-i\eta, 1/2}(-2ipr)}{r}\right), 
\label{eq:X_de} \\
X^{(I)}_{(f)} &=& y_s \frac{C_\eta e^{i\sigma_0}}{K^{(I)}(p)-2\kappa H(\eta)} 
\left(\frac{Z_N}{M_N} - \frac{Z_p}{M_p}\right) \left(2\kappa H(\eta)-\frac{2\pi}{\mu}J_0^{\mathrm{div}} \right), 
\\
X^{(I)}_{(g)} &=& -h^{(I)} \frac{4\pi Z_O}{\mu} \frac{C_\eta e^{i\sigma_0}}{K^{(I)}(p)-2\kappa H(\eta)}.
\end{eqnarray}
\end{widetext}

The integrals of Eq.~(\ref{eq:X_ab}) and Eq.~(\ref{eq:X_de}) can be rewritten in terms of the confluent 
hypergeometric functions instead of the Coulomb and Whittaker functions~\cite{ORBC}. 
In order to perform numerical integration, we simplify these integrals as
\begin{widetext}
\begin{eqnarray}
    L_{(a+b)}(p) &=& \int_0^{\infty}dr \mbox{ }e^{(-\gamma + ip)r}r^3 U(2+\kappa/\gamma, 4, 2\gamma r)\left[\frac{Z_N}{M_N} j_0
    \left(\frac{\mu}{M_N} kr\right)-\frac{Z_p}{M_p} j_0\left(\frac{\mu}{M_p} kr\right)\right] \nonumber \\
    &&\mbox{}\times \left[M(1+i\eta, 2, -2ipr)-(1+i\eta)M(2+i\eta, 3, -2ipr)\right], 
    \label{eq:L_ab} \\
    L_{(d+e)}(p) &=& \int_{r_c}^{\infty}dr\mbox{ }e^{(-\gamma + ip)r}r^3 U(2+\kappa/\gamma, 4, 2\gamma r)\left[\frac{Z_N}{M_N} 
    j_0\left(\frac{\mu}{M_N} kr\right)-\frac{Z_p}{M_p} j_0\left(\frac{\mu}{M_p} kr\right)\right] \nonumber \\
    &&\mbox{} \times \left[U(1+i\eta, 2, -2ipr) + 2(1+i\eta)U(2+i\eta, 3, -2ipr)\right] 
    \label{eq:L_de},
\end{eqnarray}
\end{widetext}
so that $X_{(a+b)}$ and $X_{(d+e)}^{(I)}$ can be reexpressed as
\begin{eqnarray}
    X_{(a+b)} &=& iy_s\frac{8 \gamma^3 p}{3}\Gamma(2+\kappa/\gamma)C_\eta e^{i\sigma_0} L_{(a+b)}, \label{eq:X_ab_2}\\
    X_{(d+e)}^{(I)} &=& y_s\frac{16 \gamma^3 p^2}{3}\Gamma(1+i\eta)\Gamma(2+\kappa/\gamma) \nonumber \\
    &&\mbox{} \times \frac{C_\eta e^{i\sigma_0}}{K^{(I)}(p)-2\kappa H(\eta)}L_{(d+e)}, \label{eq:X_de_2}
\end{eqnarray}
where $M(a, c, z)$ and $U(a, c, z)$ are the confluent hypergeometric functions of the first and second kinds, respectively.
The detailed derivation is given in the Appendix.

The cut-off $r_c$ is introduced in the integration of Eq.~(\ref{eq:L_de}) to avoid divergence.  
The divergence from its rest part and the bubble diagram in Fig.~\ref{fig:npgo}(f) are absorbed by the counter term containing the LEC
$h^{(I)}_R$, which is defined by 
\begin{equation}
    h^{(I)}_R = h^{(I)} + \frac{y_s}{Z_O}\left(\frac{Z_N}{M_N}-\frac{Z_p}{M_p}\right)\left(\frac{1}{2}J_0^{\mathrm{div}}+L_{(d+e)}^{\mathrm{div}}\right), 
    \label{h_R}
\end{equation}
where $J_0^{\mathrm{div}}$ is given in Eq.~(\ref{J0div}) and $L_{(d+e)}^{\mathrm{div}}$ is the divergence part coming from the loop 
in the diagram (d) and (e) in Fig.~\ref{fig:npgo}.
Explicitly, it reads
\begin{equation}
    L_{(d+e)}^{\mathrm{div}} = \frac{\mu}{6\pi} \int_0^{r_c} \frac{dr}{r^2} - \frac{\mu\kappa}{2\pi} \int_0^{r_c} \frac{dr}{r},
    \label{eq;rC}
\end{equation}
which has both logarithmic and linear divergences.
The dependence of the LECs and the astrophysical $S$ factor on the cut-off value of $r_c$ will be examined in the next section.

\section{Numerical results}
\label{sec:results}

The purpose of the present work is to extrapolate the astrophysical $S$ factors, or equivalently the total cross sections, of the 
$\nuclide[15]{N}(p, \gamma)\nuclide[16]{O}$ reaction to the extremely low energy regions corresponding to stellar environments 
which are hard to reach in laboratory. 
The astrophysical $S$ factor is defined by the total cross section $\sigma(E)$ that is given by
\begin{equation}
    \sigma(E) = \frac{1}{4\pi}\frac{\gamma^2 + p^2}{4p}\sum_{\mathrm{spins}}|\mathscr{M}|^2 ,
\end{equation}
in the non-relativistic limit up to the leading terms in $1/M$ expansion~\cite{AH04}, where $\mathscr{M}$ is the capture amplitude defined in Eq.~(\ref{eq:amplitude}), and $\gamma$ and $p$ are 
introduced in the previous section.
Using the measured masses of the nuclei we obtain $\gamma \approx 146$~MeV.%
\footnote{In the present work, we use $M_p = 938.272$~MeV for the proton mass, $M_N = 13968.936$~MeV for the \nuclide[15]{N} mass, and 
$M_O = 14895.081$~MeV for the \nuclide[16]{O} mass.}

The effective Lagrangian (\ref{eq:Lagrangian}) contains the following LECs: $y_s$, $y_t^{I=0,1}$, and $h^{I=0,1}$.
In addition, since we include two  $s$-wave resonance states of \nuclide[16]{O}, one with $I=0$ and the other with $I=1$, by employing ERE,
additional parameters for each resonance state are introduced as shown in Eq.~(\ref{eq:ERE}). 
However, as Eq.~(\ref{eq:prop}) shows, the inverse of the squared $y_t^{I=0,1}$ is multiplied to the corresponding di-field propagator and 
this factor is canceled out when multiplied by the vertex functions. 
Therefore, these two LECs are redundant and cannot be determined through the analysis of the present work.
The coupling constant $y_s$ determines the strength of the coupling between the ${}^3P_0$ state of the $p$-\nuclide[15]{N} system and 
the ground state of \nuclide[16]{O}.
The $h_R^{(I=0,1)}$ terms are the counter terms which absorb the divergences from the loop integrals and the values of $h_R^{(I=0,1)}$ 
depend on the cut-off $r_c$ in Eq.~(\ref{eq;rC}). 
Furthermore, throughout numerical computations, we found that the ERE for the first resonance of iso-singlet gives stable results with 
the terms up to $O(p^2)$, while $O(p^6)$ expansion is needed for the second resonance of iso-triplet.
These ERE parameters are rephrased in terms of the resonance energies ($E_{R0}$, $E_{R1}$), widths ($\Gamma_{R0}$, $\Gamma_{R1}$), 
and higher order corrections $R(E)$ for the second resonance, which introduces two parameters ($a$, $b$).
As a result, we have totally 9 parameters, namely, $y_s$, $h^{I=0,1}$, $E_{R0}$, $E_{R1}$, $\Gamma_{R0}$, $\Gamma_{R1}$, $a$, and $b$
on top of the cut-off $r_c$ value.

In this study, we work with three different data sets reported by Refs.~\cite{RR74,LIGJ10,IDBC12}.
The old data set of Ref.~\cite{RR74}, referred to as RR, contains 53 data points which cover the energy range of
$150~\mbox{keV} < E_p < 2500~\mbox{keV}$.
The 80 data points reported by Ref.~\cite{LIGJ10}, referred to as LeB, cover the energy range of $130~\mbox{keV} < E_p < 1800~\mbox{keV}$.
The most recent data reported by Ref.~\cite{IDBC12}, referred to as Imb, cover the similar energy region, 
$140~\mbox{keV} < E_p < 1800~\mbox{keV}$ with 78 data points.%
\footnote{The data set of Ref.~\cite{LIGJ10} was revised in Ref.~\cite{dGILUW13}. We use the revised data set in the present work.}
We use these data sets separately to determine our model parameters, which enables us to predict the astrophysical $S$ factor at zero energy.

\begin{figure}[t]
   \centering
   \includegraphics[width=\columnwidth]{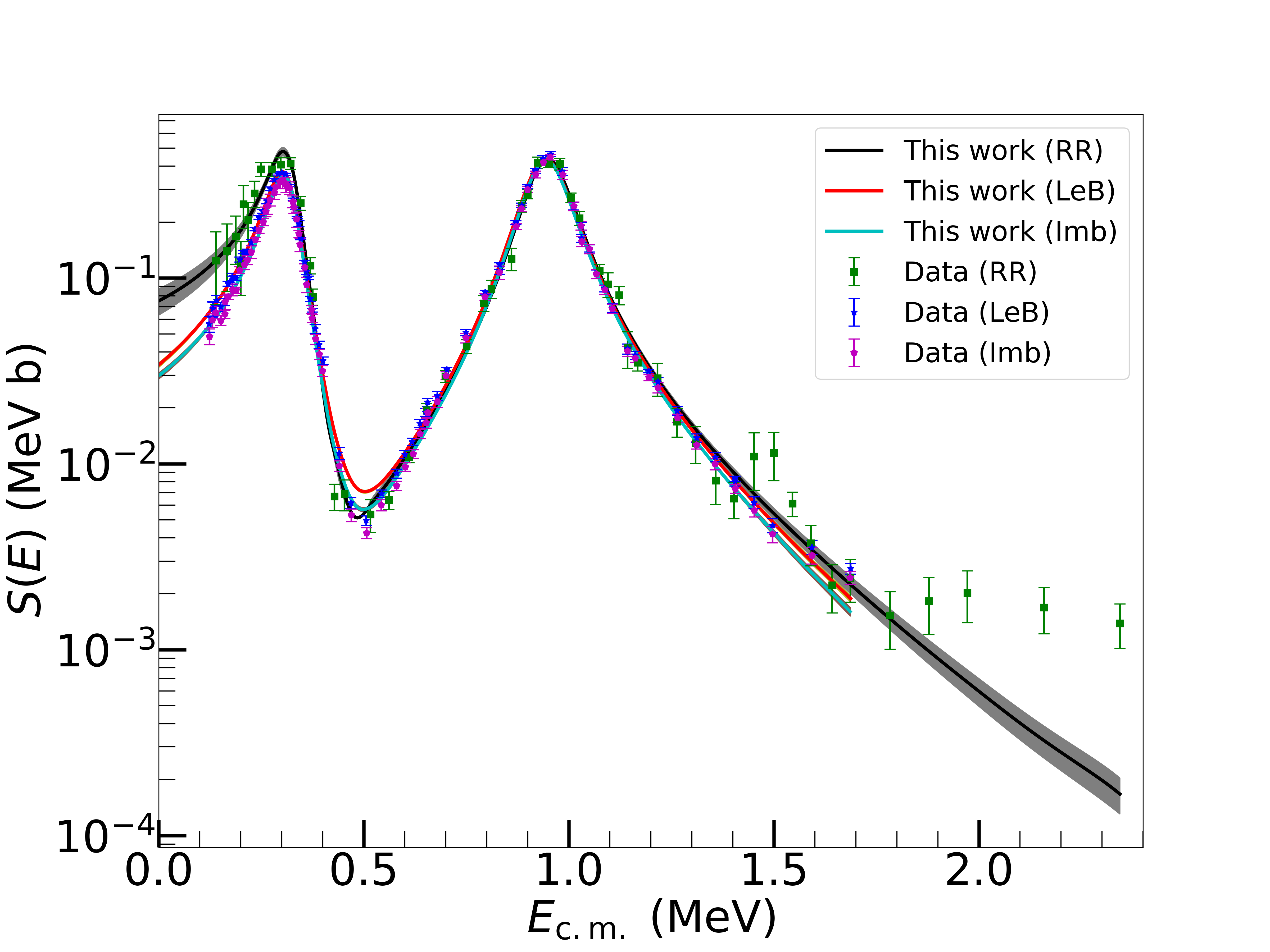}
   \caption{   \label{fig:result}
   The astrophysical $S$ factor. The parameters are fitted by the three experimental data sets from Ref.~\cite{RR74} (RR), \cite{LIGJ10} (LeB), and
   \cite{IDBC12} (Imb). 
   The bands indicate the error range obtained in the MCMC calculations.
   }
\end{figure}

\begin{figure}[t]
   \centering
   \includegraphics[width=\columnwidth]{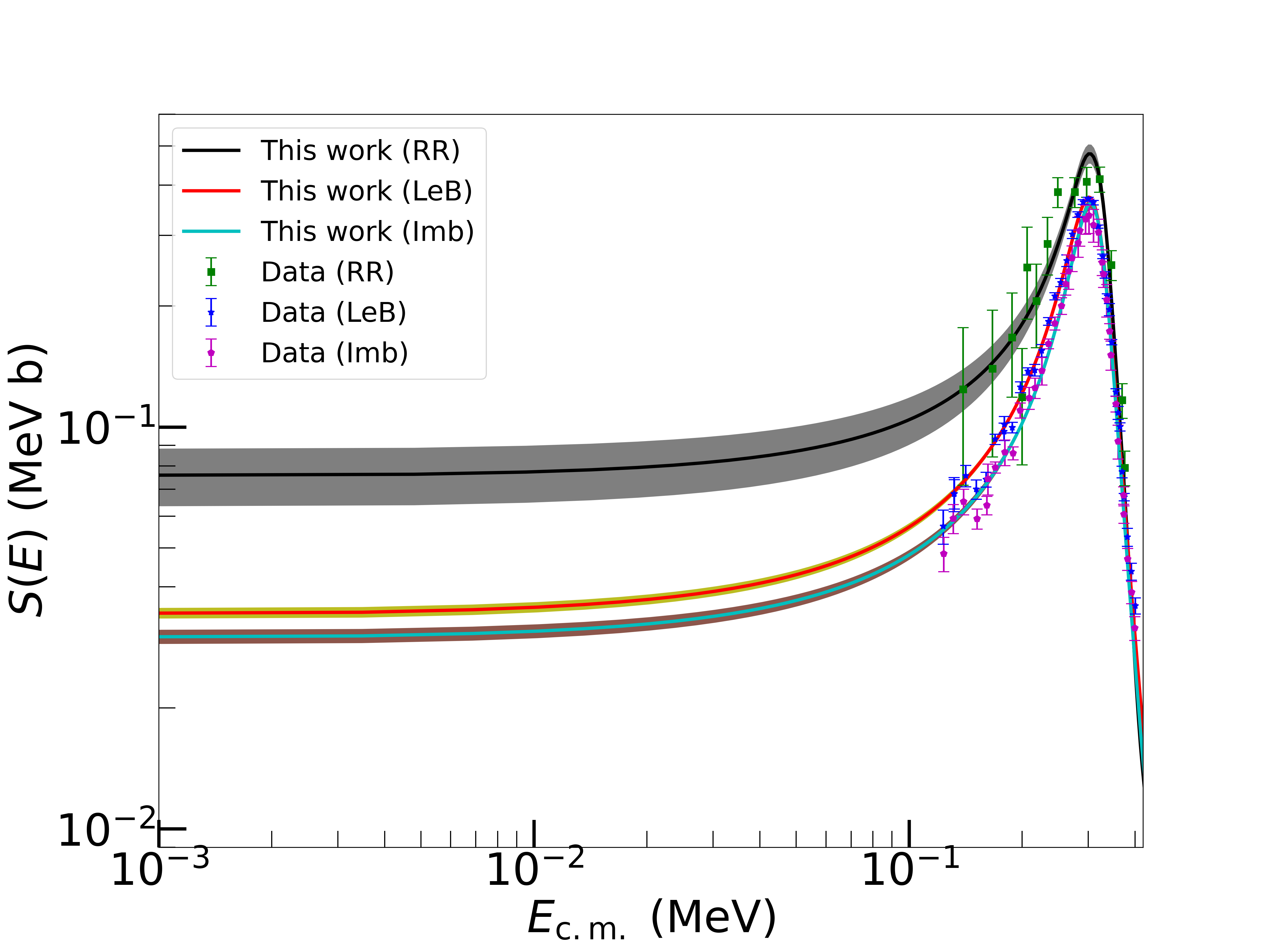}
   \caption{   \label{fig:result2}
   Same as Fig.~\ref{fig:result} but with logarithmic scales in both axes.
   }
\end{figure}

For the fitting process, we adopt the Markov chain Monte Carlo (MCMC) method~\cite{FHLG12} to minimize $\chi^2$ defined as 
$\chi^2 = \sum_i (S^{\rm Theor.}_i - S^{\rm Expt.}_i)^2 / (S^{\rm Err.}_i)^2$. 
Figure~\ref{fig:result} presents our results for the astrophysical $S$ factor of the \nuclide[15]{N}($p$,$\gamma$)\nuclide[16]{O} reaction
as a function of the c.m. energy $E$.  
In Fig.~\ref{fig:result2}, the same results are shown in logarithmic scales in both axes to highlight the extrapolated $S$ factor in the very low energy region, which 
explicitly shows that the $S$ factor becomes nearly flat as $E \to 0$.
The bands in these figures indicate the errors in the $S$ factor estimation of the MCMC calculations, which are estimated from the elements of 
the covariance matrix.

The fitted parameter values and the predicted $S(0)$ for the three data sets are given in Table~\ref{tab:results}.
This shows that the resonance parameters are consistent to each other, which is expected as the three data sets share the same structure 
arising from the two resonances.
The differences in the magnitudes, in particular, between the old~\cite{RR74} and new data~\cite{LIGJ10,IDBC12} are reflected mostly in the 
differences in the couplings $y_s$ and $h_R^{I=0}$.
This behavior is also understood as the gaps among data sets lie mostly in the very low energy region where the direct coupling ($y_s$) and 
the coupling to the iso-singlet resonance of \nuclide[16]{O} ($h_R^{I=0}$) dominate.

\begin{table*}[t]
\begin{center}
\begin{tabular}{c|ccccccccc|ccc}
\hline\hline
Data Set & $y_s$ & $h^{(I=0)}_R$ & $h^{(I=1)}_R$ & $E_{R0}$ & $\Gamma_{R0}$ & $E_{R1}$ &
$\Gamma_{R1}$ & $a$ & $b$ & $\chi^2/\mbox{datum}$ & $S(0)$ & $S(E_G)$ \\
\hline
RR~\cite{RR74} & $2.759 $ & $3.410 $ & $-29.98 $ &
  $364.6$ & $259.6$ & $967.2$ & $155.9$ & $-0.389$ & $1.158$ & 2.43 &
  $75.3 \pm 12.1$ & $80.9 \pm 12.5$
\\
\hline
LeB~\cite{LIGJ10} & $0.603$ & $0.585$ & $-31.46$ & 
  $369.3$ & $347.4$ & 962.9 & 160.0 & $-0.278$ & $1.417$ & 5.04 &
  $34.1 \pm 0.9$ & $38.2 \pm 1.0$
\\
\hline
Imb~\cite{IDBC12}  & $0.897$ & $0.978$ & $-31.05$ & 
  $359.8$ & $252.1$ & $962.6$ & $154.6$ & $-0.326$ & $1.636$ & $2.58$ &
  $29.8 \pm 1.1$ & $33.2 \pm 1.1$ 
\\
\hline\hline
\end{tabular}
\end{center}
\caption{ \label{tab:results}
The fitted values of the parameters for each data set and the derived astrophysical $S$ factor at $E=0$ and at the Gamow energy $E_G=26$~keV.
The units of $y_s$ are $10^{-2} \mbox{ MeV}^{-1/2}$ and those of $h_R^I$ are $10^{-3} \mbox{ MeV}^{1/2}$.
The energies and widths of the resonances are given in the units of $\mbox{keV}$.
The units of $a$ and $b$ are $\mbox{MeV}^{-1}$ and $\mbox{MeV}^{-2}$, respectively, and
the astrophysical $S$ factors are in the units of $\mbox{keV} \  \mbox{b}$.
For the cut-off value, we use $r_c = 1.0$~fm. }
\end{table*}

The fitted resonance parameters are $E_{R0} \approx \mbox{360 -- 370}$~keV, $\Gamma_{R0} \approx \mbox{250 -- 350}$~keV,
$E_{R1} \approx \mbox{960 -- 970}$~keV, and $\Gamma_{R1} \approx \mbox{155 -- 160}$~keV.
These should be compared with the experimentally measured values of Ref.~\cite{TWC93}, namely,
$E_{R0}^{\rm Expt.} = 312$~keV, $\Gamma_{R0}^{\rm Expt.} =  91$~keV, $E_{R1}^{\rm Expt.} = 926$~keV, 
and $\Gamma_{R1}^{\rm Expt.} = 130$~keV.
We can see that the differences between resonance parameters in the present work and the empirical values are within 
5--20\% except $\Gamma_{R0}$.
Our fitted $\Gamma_{R0}$ is more than three times the empirical value $\Gamma_{R0}^{\rm Expt.}$.
Such a difference is also seen in the $\alpha\alpha$ scattering study within EFT in Ref.~\cite{HHV08}.
Furthermore, the structures in the astrophysical $S$ factors shown in Fig.~\ref{fig:result} are reasonably reproduced by the present resonance
parameters. 
In fact, our results indicate that the resonance structure in the astrophysical $S$ factor is not sensitive to the value of $\Gamma_{R0}$,
and, therefore, constraining the value of $\Gamma_{R0}$ by this reaction is not easy.%
\footnote{The fitting was tried with forcing $\Gamma_{R0}^{\rm Expt.} =  91$~keV. In this case, although the broadness of the curve of
$S(E)$ in the first resonance region does not change, it overestimates the height of the resonance peak.}
Thus the resolution of the difference in resonance parameters deserve more detailed and comprehensive studies~\cite{HFV20}.

As in $R$-matrix analyses~\cite{Hebbard60,RR74,dGILUW13}, we find that there is a destructive interference between the two resonance 
contributions at $E \approx 0.5$~MeV, where the contributions from the two resonances overlap.
We also find that the resonance contributions dominate the cross sections and the direct emission of the photon without a resonance shown
in Figs.~\ref{fig:npgo}(a,b,c) is suppressed.

\begin{table}[t]
\centering
\begin{tabular}{c|c c c}
\hline\hline
Data Set & RR~\cite{RR74} & LeB~\cite{LIGJ10} & Imb~\cite{IDBC12} \\ \hline
$a_R^{(I=0)}$ ($10^{4}$~fm) & $-1.722$ & $-1.761$ & $-1.824$
\\ 
$\Tilde{r}^{(I=0)}$ ($10^{-3}$~fm) & $-7.053$ & $-6.993$ & $-6.746$
\\
$a_R^{(I=1)}$ (fm) & $-21.44$ & $-19.04$ & $-18.87$
\\
$\Tilde{r}^{(I=1)}$ (fm) & $-4.342$ & $-5.250$ & $-5.319$
\\
$\Tilde{P}^{(I=1)}$ ($\mathrm{fm}^3$) & $-144.07$ & $-189.5$ & $-192.6$
\\
$\Tilde{Q}^{(I=1)}$ ($\mathrm{fm}^5$) & $-246.26$ & $-339.5$ & $-345.2$
\\ \hline
$r^{(I=0)}$ (fm) & $1.457$ & $1.457$ & $1.458$ 
\\
$r^{(I=1)}$ (fm) & $-2.878$ & $-3.786$ & $-3.855$
\\
$P^{(I=1)}$ ($\mathrm{fm}^3$) & $-149.72$ & $-195.1$ & $-198.2$
\\
$Q^{(I=1)}$ ($\mathrm{fm}^5$) & $-233.28$ & $-326.5$ & $-332.2$
\\
\hline\hline
\end{tabular}
\caption{\label{tab:ERE}
Reduced effective range parameters and effective range parameters for each data set.}
\end{table}

Shown in Table~\ref{tab:ERE} are the effective range parameters and the reduced effective range parameters derived from the resonance parameters
given in Table~\ref{tab:results}.
As we have seen above, all three data sets share the similar structures in the astrophysical $S$ factor coming from the resonances, 
and it is expected that they have similar ERE parameters.
For the first resonance we take ERE up to $O(p^2)$ and for the second resonance we take up to $O(p^6)$.
Our results show that the effects from the $\kappa$ terms in the reduced effective range parameters are non-trivial, in particular, for the effective range $\Tilde{r}$.
We find that there is a big cancellation in $\Tilde{r}^{(I=0)}$ between $r^{(I=0)}$ and the $\kappa$ term, which leads to a fine tuning of
the value of $\Tilde{r}^{(I=0)}$.

\begin{table}[t]
\begin{center}
\begin{tabular}{l | c}
\hline\hline
  \multicolumn{2}{c}{ $S(0)$ ($\mathrm{keV \  b}$) } \\ \hline
  \multicolumn{2}{c}{ Data of Ref.~\cite{Hebbard60}} \\ \hline
Ref.~\cite{Hebbard60} & 32%
\footnote{This value is obtained for the proton energy of 25~keV in the laboratory frame, which gives the equivalent 
value of $S(0) \approx 26$~$\mathrm{keV \  b}$~\cite{RR74}.} \\
Ref.~\cite{Barker08b} & 35--40 \\ 
\hline
\multicolumn{2}{c}{ Data of Ref.~\cite{RR74}} \\ \hline
Ref.~\cite{RR74} & $64\pm 6$ \\
Ref.~\cite{Barker08b} & 50 \\
This work (RR) & $75.3 \pm 12.1$ \\
\hline
  \multicolumn{2}{c}{ Data of Refs.~\cite{Hebbard60,RR74}} \\ \hline
Ref.~\cite{MBBG08} & $36.0\pm 6$ \\
Ref.~\cite{HBG08} & $21.1$ \\ 
\hline
\multicolumn{2}{c}{ Data of Ref.~\cite{LIGJ10}} \\ \hline
Ref.~\cite{LIGJ10} & $39.6\pm2.6$ \\
Ref.~\cite{MLK11} &  $33.1$--$40.1$ \\
Ref.~\cite{dGILUW13}  & $40\pm 3$ \\
This work (LeB) & $34.1 \pm 0.9$ \\
\hline
\multicolumn{2}{c}{ Compiled data} \\ \hline
Ref.~\cite{XTGAOU13} & $45^{+9}_{-7}$ \\
Ref.~\cite{DD14} & $39.5$--$43.35$ \footnote{The authors of Ref.~\cite{DD14} obtained these values for $E_p = 50$--$60$ keV and accepted it as $S(0)$.}
\\ \hline
\multicolumn{2}{c}{ Data of Ref.~\cite{CMCB11}} \\ \hline
Ref.~\cite{SAO22a}  & $30.4$ \\ \hline
\multicolumn{2}{c}{ Data of Ref.~\cite{IDBC12}} \\ \hline
This work (Imb) & $29.8 \pm 1.1$ \\
\hline \hline
\end{tabular}
\end{center}
\caption{\label{tab:comp}
Estimated values of $S(0)$ for the $\nuclide[15]{N}(p, \gamma)\nuclide[16]{O}$ reaction in the units of $\mathrm{keV\ b}$.
The estimated values are categorized by the used data set(s).}
\end{table}

Since we have introduced the cut-off $r_c$ in the loop integral, the results presented in the present work are obtained by taking $r_c = 1$~fm.
In order to see the $r_c$-dependence of our results, we varied the values of $r_c$ from 0.1~fm to 1.0~fm to confirm that the $r_c$-dependence of 
our results on the astrophysical $S$ factor are very weak although the coupling strengths would change.

\begin{figure*}[t]
   \centering
   \includegraphics[width=\textwidth]{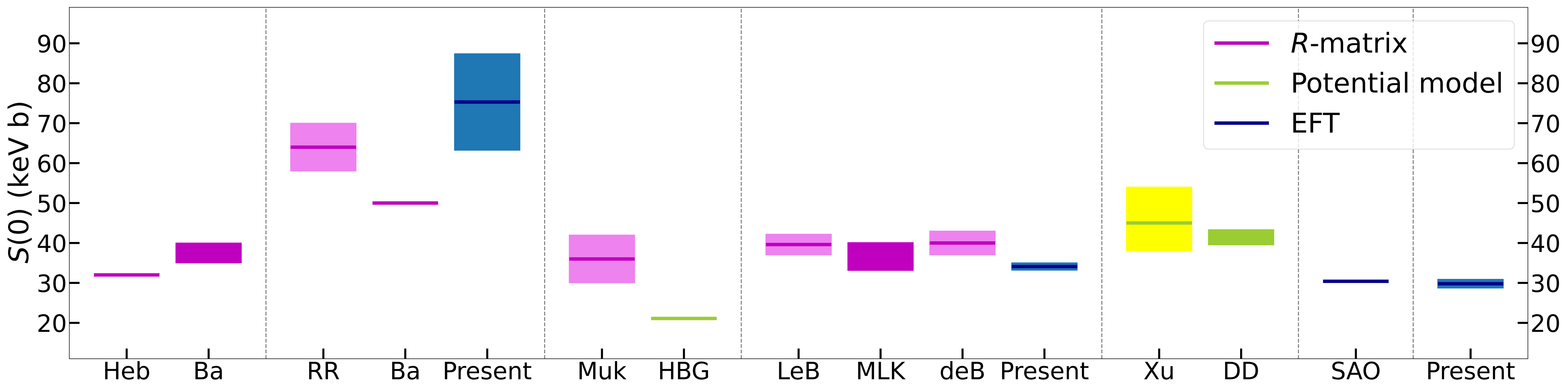}
   \caption{   \label{fig:comp}
   The obtained $S(0)$ values of the present work compared with the values in the literature.
   The abbreviated references are as follows: Heb~\cite{Hebbard60}, Ba~\cite{Barker08b}, RR~\cite{RR74}, Muk~\cite{MBBG08}, HBG~\cite{HBG08}, 
   LeB~\cite{LIGJ10}, MLK~\cite{MLK11}, deB~\cite{dGILUW13}, Xu~\cite{XTGAOU13}, DD~\cite{DD14}, Caci~\cite{CMCB11}, SAO~\cite{SAO22a}, and Imb~\cite{IDBC12}.
   The references in the upper part indicate the sources of the experimental data and the references in the lower part indicate the sources of the
   calculated results.
   }
\end{figure*}

In Table~\ref{tab:comp}, we list the previous estimates of the astrophysical $S$ factor at zero energy $S(0)$
from the $R$-matrix analysis~\cite{RR74, Hebbard60, LIGJ10, Barker08b, MLK11, MBBG08, dGILUW13} or potential model calculations~\cite{HBG08, XTGAOU13, DD14}.
Those values of $S(0)$ are scattered from 22~$\mathrm{keV\  b}$ to 64~$\mathrm{keV \  b}$, depending on the data used for the fit.
We first note that our estimate on the central value of the astrophysical $S(0)$ using the data of Ref.~\cite{RR74} is somehow larger than 
the values of the $R$ matrix analyses of Refs.~\cite{RR74,Barker08b} but the large error ranges overlap each other.
However, these estimates differ from the $S(0)$ value extracted from the post-NACRE data of Refs.~\cite{LIGJ10,IDBC12},
which give new values around 30~$\mathrm{keV \  b}$. 
Our results are also close to the values extracted from the data sets up to the energy region of the first resonance reported 
in Refs.~\cite{XTGAOU13,DD14,SAO22a}.
The comparison of our results for $S(0)$ with the values in the literature are visualized as well in Fig.~\ref{fig:comp}.
The results presented in Table~\ref{tab:comp} also shows that the $S$ factor at the Gamow energy is about 10\% larger than $S(0)$.

\section{Summary and discussion}
\label{sec:summary}

In the present work, we studied the radiative proton capture process, \nuclide[15]{N}($p$,$\gamma$)\nuclide[16]{O}, that has 
an important role in the CNO cycle by connecting the CN cycle and the NO cycle.
Our study was performed by employing the EFT formalism, and, to our knowledge, there was no application of EFT approach for
the investigation of this reaction.
Three open channels of $\alpha$-\nuclide[12]{C}, $\alpha$-\nuclide[12]{C}$^*$, and $p$-\nuclide[15]{N} can be involved to describe 
the resonant states of \nuclide[16]{O}, and the experimental data of the \nuclide[15]{N}($p$,$\gamma$)\nuclide[16]{O} reaction
clearly show the structure of the two resonant $1^-$ states of $\nuclide[16]{O}^*$.
One of them is the isospin singlet state and the other is the isospin triplet state, which have the excitation energies of 12.45~MeV and
13.09~MeV, respectively.

In order to describe this reaction, we employ a minimum set of effective Lagrangian with the terms for the two $1^-$ states
with $I=0$ and $I=1$ in the $p$-\nuclide[15]{N} channel. 
The resonances are described by the di-fields and the structure of their propagators are rephrased in terms of ERE and 
by taking the Breit-Wigner form.
The parameters introduced in the low-energy effective Lagrangian are fitted to the avilable experimental data of the astrophysical $S$ factor of
the \nuclide[15]{N}($p$,$\gamma$)\nuclide[16]{O} reaction in the range of $E_p= 130$~keV to 2500~keV. 
In the present analysis, we employ the experimental data reported in Refs.~\cite{RR74,LIGJ10,IDBC12} and the fitting process was performed
for each data set using the MCMC method.
Then the astrophysical $S$ factors at $E=0$ and $E=E_G$ are extrapolated and compared with the estimates in the literature.
Our results show that the value of $S(0)$ is in the range of 30--35~$\mathrm{keV \  b}$ based on the data of Refs.~\cite{LIGJ10,IDBC12}, 
which agrees with other estimates based on the post-NACRE data.
We also found that the values of $S(E_G)$ are about 10\% larger than those of $S(0)$.

The present work shows that the EFT approach can be applied to the study of \nuclide[15]{N}($p$,$\gamma$)\nuclide[16]{O}.
However, the extracted resonance parameters are puzzling as they show some discrepancies with the empirical values,
in particular for the first resonance state although the resonance structures in the astrophysical $S$ factor are well reproduced.
The overestimate of a resonance width was also found in $\alpha\alpha$ scattering study within EFT,
but the experimental value of a width of \nuclide[16]{O} could be well reproduced in the study of the $S$ factor of 
\nuclide[12]{C}($\alpha$,$\gamma$)\nuclide[16]{O} within EFT~\cite{Ando18}. 
This deserves more detailed and rigorous investigations including the effects of other open channels.

\acknowledgments

\newblock
We are grateful to T.-S. H. Lee for fruitful discussions.
This work was supported by the National Research Foundation of Korea (NRF) under Grants No.~NRF-2019R1F1A1040362, 
No.~NRF-2022R1F1A1070060, No.~NRF-2020R1A2C1007597, and No.~NRF-2018R1A6A1A06024970 (Basic Science Research Program).

\appendix*

\section{Loop integrals}

In this Appendix, we derive the integrals of Eqs.~(\ref{eq:L_ab}) and (\ref{eq:L_de}) explicitly by using identities of the Coulomb functions, 
Whittaker functions, and the confluent hypergeometric functions~\cite{ORBC}. 
The Coulomb function $F_l(\eta, \rho) $ is written as
\begin{equation}
    F_l(\eta, \rho) = C_l(\eta)\rho^{l+1}e^{i\rho}M(l+1+i\eta, 2l+2, -2i\rho), 
    \label{A:coulomb}
\end{equation}
where $M(a, c, z)$ is the confluent hypergeometric function of the first kind or the Kummer's function. 
The factor of $C_l(\eta)$ is defined as
\begin{equation}
    C_l(\eta) = \frac{2^l}{(2l+1)!}|\Gamma(l+1+i\eta)|e^{-\frac{\pi}{2}\eta},
\end{equation}
which becomes the Gamow factor for $l=0$, namely, $C_0(\eta)$ becomes $C_\eta$ of Eq.~(\ref{eq:Gamow}).
On the other hand, the Whittaker function $W_{\kappa, \mu}(z)$ is written as
\begin{equation}
 W_{\kappa, \mu}(z) = \textstyle e^{-\frac{z}{2}}z^{\mu+\frac{1}{2}} U (\mu - \kappa + \frac{1}{2}, 1+2\mu, z ), \label{A:whittaker} 
\end{equation}
where $U(a, c, z)$ is the confluent hypergeometric function of the second kind or the Tricomi's function.

Using Eqs.~(\ref{A:coulomb}) and (\ref{A:whittaker}), it is straightforward to get
\begin{eqnarray}
    W_{-\kappa/\gamma, 3/2}(2\gamma r) &=& 4\gamma^2 r^2 e^{-\gamma r}U(2+\kappa/\gamma, 4, 2\gamma r), \\
    \frac{W_{-i\eta, 1/2}(-2ipr)}{r} &=& -2ip e^{ipr}U(1+i\eta, 2, -2ipr), \\
        \frac{F_0(\eta, pr)}{pr} &=& C_\eta e^{ipr}M(1+i\eta, 2, -2ipr).
\end{eqnarray}
As the loop integrals of Eqs.~(\ref{eq:X_ab}) and (\ref{eq:X_de}) contain derivatives of the above functions, we 
use the identities of the derivative of the confluent hypergeometric functions as
\begin{eqnarray}
    \frac{d}{dz}M(a, c, z) &=& \frac{a}{c}M(a+1, c+1, z), \\
    \frac{d}{dz}U(a, c, z) &=& -aU(a+1, c+1, z)
\end{eqnarray}
to obtain the formulas of Eqs.~(\ref{eq:X_ab_2}) and (\ref{eq:X_de_2}).
This finally leads to the integrals of Eqs.~(\ref{eq:X_ab}) and (\ref{eq:X_de}) in the simplified form of
\begin{widetext}
\begin{eqnarray}
    && \int_0^{\infty}dr r W_{-\kappa/\gamma, 3/2}(2\gamma r)\left[\frac{Z_N}{M_N}j_0\left(\frac{\mu}{M_N}kr\right)-\frac{Z_p}{M_p}
    j_0\left(\frac{\mu}{M_p}kr\right)\right]\frac{\partial}{\partial r}\left(\frac{F_0(\eta, pr)}{pr}\right) 
    \nonumber \\
    & =&  4\gamma^2 ip C_\eta \int_0^{\infty}dr \mbox{ }e^{(-\gamma + ip)r}r^3 U(2+\kappa/\gamma, 4, 2\gamma r)\left[\frac{Z_N}{M_N} 
    j_0\left(\frac{\mu}{M_N} kr\right)-\frac{Z_p}{M_p} j_0\left(\frac{\mu}{M_p} kr\right)\right] 
    \nonumber \\ && \qquad \mbox{}  
    \times \left[M(1+i\eta, 2, -2ipr)-(1+i\eta)M(2+i\eta, 3, -2ipr)\right], 
\label{A:L_ab} 
\end{eqnarray}
and
\begin{eqnarray}
    && \int_{0}^\infty dr r W_{-\kappa/\gamma, 3/2}(2\gamma r)\left[\frac{Z_N}{M_N} j_0\left(\frac{\mu}{M_N} kr\right)-\frac{Z_p}{M_p} 
    j_0\left(\frac{\mu}{M_p} kr\right)\right]\frac{\partial}{\partial r}\left(\frac{W_{-i\eta, 1/2}(-2ipr)}{r}\right) \nonumber \\
    &=&  8\gamma^2 p^2 \int_{0}^{\infty}dr\mbox{ }e^{(-\gamma + ip)r}r^3 U(2+\kappa/\gamma, 4, 2\gamma r)\left[\frac{Z_N}{M_N} 
    j_0\left(\frac{\mu}{M_N} kr\right)-\frac{Z_p}{M_p} j_0\left(\frac{\mu}{M_p} kr\right)\right] 
    \nonumber \\ && \qquad \mbox{} 
    \times \left[U(1+i\eta, 2, -2ipr) + 2(1+i\eta)U(2+i\eta, 3, -2ipr)\right]. \label{A:L_de}
\end{eqnarray}
\end{widetext}
Embedding the cut-off $r_c$ on the integration range of Eq.~(\ref{A:L_de}) gives Eqs.~(\ref{eq:X_ab_2}) and (\ref{eq:X_de_2}).

\end{document}